\documentclass[12pt]{iopart}
\usepackage{graphicx}

\expandafter\let\csname equation*\endcsname\relax
\expandafter\let\csname endequation*\endcsname\relax
\usepackage{amsmath,amsfonts,amssymb}
\usepackage[caption=false]{subfig}
\usepackage[dvipsnames,usenames]{xcolor}
\usepackage{ulem}
\usepackage{fancyhdr}

\definecolor{linkcolor}{rgb}{0.0,0.3,0.5}
\usepackage[
unicode, 
colorlinks=true,
linkcolor=linkcolor,
citecolor=linkcolor,
filecolor=linkcolor,
urlcolor=linkcolor,
hyperfootnotes=true,
]{hyperref}% add hypertext capabilities
\usepackage{epsfig}
\usepackage{graphicx}
\usepackage{bm}

\usepackage{float}
\usepackage{leftidx}
\usepackage{empheq} 
\usepackage{color}

\usepackage{censor}
\StopCensoring

\usepackage[numbers,square,sort&compress]{natbib}

\usepackage{array}
\newcolumntype{C}[1]{>{\vspace{5pt}\centering\let\newline\\\arraybackslash\hspace{0pt}\vspace{5pt}}m{#1}}

\newcommand{\mpl}{M_\mathrm{Pl}}

\newcommand{\psibh}{\psi_\mathrm{BH}}
\newcommand{\parpar}{\partial^j\partial_j}
\newcommand{\cttk}{CTTK}

\newcommand{\grchombo}{\textsc{grchombo}}

\newcommand{\eqsref}[1]{(\ref{#1})}
\renewcommand{\eqref}[1]{Eqn. (\ref{#1})}

\makeatletter
\newcommand{\mainmatter}{%
  \let\@fnsymbol\@arabic
  \def\@makefnmark{\textsuperscript{\arabic{footnote}}}%
}
\makeatother

\graphicspath{{figures/}}

\begin{document}

\normalem
% {\hfill KCL-PH-TH/2022-39}
\title{\cttk: A new method to solve the initial data constraints in numerical relativity}

\author{Josu C. Aurrekoetxea}
\ead{josu.aurrekoetxea@physics.ox.ac.uk}
\address{Astrophysics, University of Oxford, Denys Wilkinson Building, Keble Road, Oxford OX1 3RH, United Kingdom}
\author{Katy Clough}
\ead{k.clough@qmul.ac.uk}
\address{School of Mathematical 
Sciences, Queen Mary University of London, Mile End Road, London E1 4NS, 
United Kingdom}
\author{Eugene A. Lim}
\ead{eugene.a.lim@gmail.com}
\address{Theoretical Particle Physics and Cosmology Group, Physics 
Department, King's College London, Strand, London WC2R 2LS, United Kingdom}

\begin{abstract}

In numerical relativity simulations with non-trivial matter configurations, one must solve the Hamiltonian and momentum constraints of the ADM formulation for the metric variables in the initial data. We introduce a new scheme based on the standard Conformal Transverse-Traceless (CTT) decomposition, in which instead of solving the Hamiltonian constraint as a 2nd order elliptic equation for a choice of mean curvature $K$, we solve an \emph{algebraic} equation for $K$ for a choice of conformal factor.  By doing so, we evade the existence and uniqueness problem of solutions of the Hamiltonian constraint without using the usual conformal rescaling of the source terms. This is particularly important when the sources are fundamental fields, as reconstructing the fields' configurations from the rescaled quantities is potentially problematic. Using an iterative multigrid solver, we show that this method provides rapid convergent solutions for several initial conditions that have not yet been studied in numerical relativity; namely (i) periodic inhomogeneous spacetimes with large random Gaussian scalar field perturbations and (ii) asymptotically flat black hole spacetimes with rotating scalar clouds.

\end{abstract}

\maketitle
\mainmatter

\section{Introduction} \label{sect:intro}

\noindent
In the Arnowitt-Deser-Misner (ADM) formulation of the Einstein equations \cite{Arnowitt:1962hi} a full specification of the initial conditions requires values for the various metric and matter components to be provided across a 3D spatial hyperslice of the 3+1D spacetime with normal vector $n^\mu$.  In the metric sector, one must provide six values for the spatial metric $\gamma_{ij}$ and six for the extrinsic curvature $K_{ij}$, plus four for the lapse $\alpha$ and shift $\beta^i$, where in coordinates adapted to the slicing
\begin{equation}
	ds^2 = - (\alpha^2 - \beta^i \beta_i) dt^2 + 2 \beta_i dx^i dt + \gamma_{ij} dx^i dx^j ,
\end{equation}
and
\begin{equation}
   K_{ij} = -\frac{1}{2} \mathcal{L}_n \gamma_{ij} .
\end{equation}

One must also specify the initial quantities on the 3D hypersurface in the matter sector. For example, in the case of ideal fluids, these would be the rest frame pressure and energy density, which relate directly to elements of the stress energy tensor $T^{\mu\nu}$. In other cases, the fundamental quantities are fields from which the stress energy tensor is derived in a non trivial way, for example, in the case of a scalar field, one must specify the field $\phi$ and its conjugate momentum $\Pi$, since $T_{\mu\nu}$ is a function of ($\phi$, $\Pi$) and derivatives.

The matter and metric sectors cannot be freely and independently specified, but must satisfy four constraints, the Hamiltonian constraint
\begin{equation}\label{eq:Ham_eq1}
\mathcal{H}\equiv \leftidx{^{(3)}}R+K^2-K_{ij}K^{ij}-16\pi\rho =0,
\end{equation}
and the momentum constraints
\begin{equation}\label{eq:Mom_eq1}
\mathcal{M}^i\equiv D_{j}\left(K^{ij}-\gamma^{ij}K\right)
-8\pi S^i =0,
\end{equation}
where we have set $G=c=1$. Here we define the matter quantities $\rho = n^\mu n^\nu T_{\mu\nu}$ and $S^i= -n^\mu (\gamma^{\nu i} + n^\nu n^i) T_{\mu\nu}$ as projections of the stress energy tensor.

Usually a physical problem requires that matter fields and their boundary conditions are specified, and one must solve the four constraints for the corresponding metric variables\footnote{Although it is sometimes useful to take the reverse approach, specifying the metric for a given spacetime, and reverse engineering the matter configuration to which it corresponds, see e.g. \cite{Mertens:2015ttp}. Depending on the simplicity of the matter sector, this may or may not be easier than the forward problem - in particular for fundamental fields the configuration that generates a particular curvature is not necessarily unique.}. 
Since the four constraints alone are insufficient to specify all of the metric variables listed above, some components must be chosen either arbitrarily or (if possible) using symmetries and physical intuition. 

Of the 16 metric components that must be specified (counting for completeness the lapse and shift), eight represent gauge choices, four are related to the two dynamical degrees of freedom characterising a gravitational field in general relativity and their time derivatives, and the remaining four are set by the constraints.  In most, if not all cases, there exists more than one possible metric solution to the specified matter configuration and boundary conditions -- for example, Kerr and Schwarzschild are both asymptotically flat vacuum solutions, but are physically different.

The canonical approach that is taken to ameliorate this non-uniqueness is to use a conformal decomposition of the quantities to set more fundamental quantities that can (hopefully) be chosen in a more well-motivated way. As described in \ref{sec-method}, the metric is usually decomposed into a conformal metric $\bar{\gamma}_{ij}$ with unit determinant and a conformal factor $\psi$, whilst the extrinsic curvature is decomposed into the mean curvature ($K = K^i_i$), its transverse-traceless part ($\bar{A}^{ij}_\mathrm{TT}$) and longitudinal part (described by a vector potential $W^i$).

Several schemes exist for choosing which variables to set and which to solve for; the most well-known of which are the Conformal Transverse Traceless (CTT) and Extended Conformal Thin Sandwich (XCTS) approaches \cite{York:1998hy,Pfeiffer:2002iy}. 
In the CTT method, the conformally related metric and the mean curvature $K$ are freely chosen, and the Hamiltonian and momentum constraints are solved for the conformal factor $\psi$ and the vector potential $W^i$ respectively. The latter XCTS approach is more suited to obtaining equilibrium initial data, and involves solving the momentum constraint for the shift vector $\beta^i$ in place of $W^i$, and an additional Poisson equation for the lapse $\alpha$, in order to obtain a (user-specified) time evolution in the conformal metric $\partial_t \bar{\gamma}_{ij}$ and mean curvature $\partial_t K$. The solution of the Hamiltonian constraint for the conformal factor $\psi$ is essentially the same as in the CTT case, and requires a choice of $K$ and the conformal metric $\bar{\gamma}_{ij}$.
(See \cite{Baumgarte:2010ndz, Gourgoulhon:2007ue, Alcubierre:2008co, Tichy:2016vmv} for reviews and other implementations \cite{East:2012zn,Assumpcao:2021fhq}).

In this work we propose a different approach to solving the constraints, in which instead of choosing $K$ and solving for $\psi$,  one does the reverse - choosing an initial profile for the conformal factor $\psi$ and solving for the trace of the extrinsic curvature $K$. This results in a simple algebraic equation for $K$ in place of the elliptic Hamiltonian constraint for $\psi$, which avoids many of the issues associated with uniqueness and existence of solutions \cite{OMurchadha:1973byk,Pfeiffer:2005jf,Baumgarte:2006ug,Walsh:2006au,Cordero-Carrion:2008grk}, albeit at the cost of a somewhat more complicated system for the momentum constraints that is unavoidably coupled to the Hamiltonian constraint. A physical interpretation of our method is that it gives the locally Friedmann-Lemaître-Robertson-Walker (FLRW) slicing of a cosmological spacetime, with $K^2 \sim \rho$, while the proper lengths on the initial spatial hypersurface are specified via the choice of $\psi$, which plays the role of the initial (local) scale factor. We have found that in several physical problems of interest, this makes the solution simpler to obtain, with more robust and consistent convergence compared to other methods. Moreover, in many cases (in particular, fundamental fields with large densities) it gives greater control over the initial conditions as proper lengths are fixed on the initial hypersurface. 
We call it the ``\cttk{} method'' to distinguish it from the traditional CTT approach in which one solves primarily for $\psi$.

This paper is structured as follows: in Sec. \ref{sec-background} we explain the issues associated with solving the constraint equations, how these apply for fundamental fields, and the advantages of our approach. In Sec. \ref{sec-ourmethod} we set out our \cttk{} approach in full, as an adaptation of the CTT method.  In Sec. \ref{sec-results} we demonstrate that our method results in convergence of the solutions in several cases of physical interest, including cosmological spacetimes with large amplitude perturbations and black holes with rotating scalar clouds. We conclude and discuss extensions in Sec. \ref{sec-conclude}. For readers not familiar with the standard CTT approach, we review its key features in \ref{sec-method}.

\section{Uniqueness and existence of solutions}
\label{sec-background}

In the CTT (and \cttk{}) approaches, the conformal quantities to be specified $\bar{\gamma}_{ij}$, $\psi$, $K$, $W^i$ and $\bar{A}^{ij}_\mathrm{TT}$ must satisfy relations of the form
\begin{align}
8\bar{D}^2\psi - \psi \bar{R} - \frac{2}{3}\psi^5 K^2 +\psi^{-7} \bar{A}_{ij}\bar{A}^{ij} &= -16\pi\psi^5\rho, \label{eq:Ham_constraint}\\
\left(\bar{\Delta}_\mathrm{L}W\right)^i - \frac{2}{3}\psi^6\bar{\gamma}^{ij}\bar{D}_j K &= 8\pi\psi^{10} S^i ,\label{eq:Mom_constraint}\\
\bar{D}_j\bar{A}^{ij}_\mathrm{TT} &=0 ~.
\end{align}
where $\bar{\Delta}_\mathrm{L}$ is the vector Laplacian (see \eqref{eq:vec_lap_eq}), $\bar{R}$ and $\bar{D}$ are the Ricci scalar and the covariant derivative associated with the conformal metric $\bar{\gamma}_{ij}$, and $\bar{A}^{ij} = \bar{A}^{ij}_\mathrm{TT} +  \bar{D}^i W^j + \bar{D}^j W^i - \frac{2}{3}\bar{\gamma}^{ij} \bar{D}_k W^k$. These coupled, non linear elliptic equations give rise to a number of pitfalls related to uniqueness and existence of solutions, which we will discuss in detail below.

The basic problem is that, as noted in the Introduction, there are multiple solutions to this system, and so we need to impose conditions to narrow it down to a single one that is as close to the physical problem of interest as possible. If we are not careful in how we choose the free quantities, we will over-restrict the system, and as a result no solutions will exist to the constraints for the chosen sources and boundary conditions. This is the problem of {\it existence}. On the other hand, if we do not restrict sufficiently, we may end up with non-unique solutions where, even if they are physically equivalent, our numerical solver will have difficulty consistently converging to a single one. This is the problem of {\it uniqueness}, which comes up most notably in systems with periodic boundary conditions \cite{Garfinkle:2020iup}.

A tried-and-tested numerical method for the solution of non-linear elliptical differential equations is to linearise the equations, and iterate from a trial solution until they converge to the final non-linear solution. If the trial solution is within the domain of convergence (usually true if the trial is close to the true solution), the method should converge.
For example, consider the Hamiltonian constraint \eqref{eq:Ham_constraint} in the simplified case with $\bar{A}^{ij}=0$ and $\bar{\gamma}_{ij}=\delta_{ij}$,
\begin{equation}
8\parpar \psi - \frac{2}{3}\psi^5 K^2 = -16\pi\psi^5\rho~,\label{eqn:Ham_reduced}
\end{equation}
in the domain $D$ with boundary $\partial D$ \cite{York:1978gql}.
This equation is non-linear in $\psi$, therefore to solve it numerically we linearise as $\psi\rightarrow \psi_0 + u$, such that \eqref{eqn:Ham_reduced} becomes
\begin{equation}
\parpar u - \left(\frac{5}{12}\psi_0^4 K^2 - 10\pi \psi_0^4\rho\right)u \equiv \left[\parpar -q({\bf x})\right] u = \sigma(\psi_0)~. \label{eqn:Ham_linear}
\end{equation}
Here the source $\sigma(\psi_0) = -2\pi\psi_0^5\rho-\parpar\psi_0 + (1/12)\psi_0^5 K^2$, which is independent of $u$, measures the difference between the trial solution $\psi_0$ and the true solution. 

Consider the case with Dirichlet boundary conditions where $u$ is specified on $\partial D$. Any trial solution $\psi_0$ must always satisfy the required boundary conditions  and hence $u(\partial D)=0$ is homogeneous by construction. If a unique solution $u$ can be found for the linearised elliptic equation, then we correct our trial solution $\psi_0 \rightarrow \psi_0 + u$, and iterate until we converge to a final solution where $u=0$. Suppose $\psi_0=\psi_*$ is a solution, then $\sigma(\psi_*)=0$ by construction, and \eqref{eqn:Ham_linear} reduces to a homogeneous elliptic PDE.
This is the relevant case for most asymptotically flat spacetimes, such as in numerical relativity evolutions of compact binaries \cite{Suh:2016ctx}.
If $q({\bf x})>0$ in $D\cup\partial D$, then the maximal principle implies that the only solution is the trivial solution $u=0$ and hence $\psi_*$ exists and is unique. % with $u(\partial D)=0$.
Conversely, in the case in which $ q({\bf x}) < 0$, the linear homogeneous equation becomes $[\parpar  + \vert q({\bf x})\vert]u= 0$, then this cannot be guaranteed, implying that $\psi_*$ is not necessarily a unique solution \cite{York:1978gql,1979Smarr,Baumgarte:2006ug,Walsh:2006au}. Notice that the sign of $q({\bf x})$ depends on the sign of the exponent of $\psi$ in the RHS of \eqref{eqn:Ham_reduced}, so this ``wrong sign'' can be corrected by a suitable rescaling of $\rho$ with additional powers of $\psi$, as we will see later\footnote{
A useful framework to understand this is to recast the linear problem using the Fredholm Alternative Theorem. Consider a self-adjoint linear operator ${\cal L}$ acting on a vector space $u$ with source $\sigma$, ${\cal L} u = \sigma$, with some boundary conditions. One of the following statement must be true: either the homogeneous problem ${\cal L}u=0$ has a non-trivial solution $u\neq 0$ or
the inhomogeneous problem ${\cal L}u=\sigma$ has a unique solution.  
In addition, if no unique solution exists,  a solution  exists if and only if the source is orthogonal to the Kernel of the operator $\int_D dV \tilde{u}\sigma=0$, where $\tilde u$ is the kernel of ${\cal L}$ i.e.  ${\cal L}\tilde u =0$.}.

Periodic boundary conditions represent a common scenario in many numerical relativity applications in cosmology when trying to simulate a small representative patch of the whole spacetime, effectively imposing regularity on the periodic scale \cite{Bentivegna:2013xna,East:2015ggf,Clough:2016ymm,Clough:2017efm,Aurrekoetxea:2019fhr,Joana:2020rxm,Yoo:2018pda,Yoo:2020lmg,deJong:2021bbo,Garfinkle:2008ei,Cook:2020oaj,Ijjas:2020dws,Ijjas:2021wml,Ijjas:2021zyf,Widdicombe:2018oeo,Giblin:2019nuv,Kou:2019bbc,Kou:2021bij,Joana:2022uwc,Yoo:2012jz,Yoo:2013yea,Yoo:2014boa,Giblin:2015vwq,Bentivegna:2015flc,Bentivegna:2016fls,Bentivegna:2018koh,Giblin:2019pql}\footnote{See \cite{Corman:2022rqo} for an alternative choice of boundary conditions.}.
In these cases, we will often find ourselves in the situation in which the solutions are not unique, for example operators like $\parpar$ are not invertible on a torus. Therefore, in many cases the numerical algorithm will fail to converge and, even when it does, it is not clear that the solution is the desired physical one. As an example, a different constant value of $\psi$ will result in different gradient energies for the same inhomogeneous scalar field (coordinate) configuration $\phi(x)$, since $\rho_\mathrm{grad}\sim \psi^{-4} (\partial_i \phi)^2$.
In addition, we have a set of four integrability conditions that must be satisfied -- one for Hamiltonian constraint
\begin{align}\label{eq:K_integrability}
\mathcal{I}_\mathcal{H}\equiv &\int_D dV \left(  \frac{2}{3}\psi^5 K^2 -16\pi\psi^5\rho\right) =0~,
\end{align}
and three for the momentum constraints\footnote{Note that we have assumed that the ansatz chosen for the vector Laplacian must satisfy $\int_D \left(\bar{\Delta}_\mathrm{L}W\right)_idV=0$, in our case given by \eqref{eq:vec_lap_eq_flat}.} given by 
\begin{align}\label{eq:S_integrability}
    \mathcal{I}_{\mathcal{M}_i} \equiv & \int_D dV \left(  \frac{2}{3}\psi^6\partial_i K + 8\pi \psi^6 S_i\right) =0~,
\end{align}
which come from integrating Eqns. \eqsref{eq:Ham_constraint} and \eqsref{eq:Mom_constraint} over the domain (note that $\int_D \partial^i\partial_i\psi= 0$ and we are still considering the simplified case where $\bar{A}^{ij}=0$ and $\bar{\gamma}_{ij}=\delta_{ij}$).
Suppose that $K$ is a constant, the momentum constraints impose on $S_i$ the restriction that there is no net momentum in any direction over the spatial slice, $\int_D dV \psi^6 S_i = 0$. This physical restriction is related to the assumption of conformal flatness, which would need to be relaxed where other solutions are of interest. When both $K$ and $\psi$ are not constant, as it could be the case for cosmological spacetimes containing black holes,  \eqref{eq:S_integrability} imposes an extra condition on the first term. These restrictions on the momentum constraint sources are discussed in greater detail in \cite{Garfinkle:2020iup}.

In the usual CTT method, a common approach is to uncouple the momentum constraints from the Hamiltonian constraint by choosing a constant value of the mean curvature $K=\bar{K}$. Naively, one could try to choose this constant value such the ``right sign'' is recovered in \eqref{eqn:Ham_linear}
\begin{equation}
    q({\bf x})\equiv \frac{5}{12}\psi_0^4\left( \bar{K}^2 - 24\pi \rho\right) ~> 0
\end{equation}
However, for asymptotically flat spacetimes the value of $\bar{K}$ is imposed by the boundary conditions to be zero. For periodic boundary conditions the constant choice of $\bar{K}$ is not free and must satisfy the integrability condition in \eqref{eq:K_integrability}. In effect, this imposes that the constant be approximately set to a volume averaged value of the energy density $\bar{\rho}$. The condition for $q({\bf x})$ can then be rewritten as 
\begin{equation}
q({\bf x}) \sim \bar{\rho} - \rho ~>0 ~.
\end{equation}
We see that this cannot be satisfied everywhere on the hyperslice\footnote{Note that the condition from the linearisation will be different if $\rho=\rho(\psi)$, as it is the case for the gradient energy density of a scalar field, with $\rho\sim \psi^{-4}(\partial_i \phi)^2$ and one can choose a constant $\bar{K}$ that satisfies the integratibility condition \cite{Clough:2016ymm,Clough:2017efm,Aurrekoetxea:2019fhr,Yoo:2018pda,Yoo:2020lmg,deJong:2021bbo}.} - by construction there will be regions where the local value of $\rho$ is smaller than its average $\bar{\rho}$ and thus $q({\bf x})>0$, but also regions where $q({\bf x})<0$ and so often no unique solution can be found.

A standard trick to evade this problem in the case of fluids is to rescale the energy density \cite{York:1978gql} by
\begin{equation}
\rho \rightarrow \tilde{\rho}\psi^{-n}~, \label{eqn:rho_rescale}
\end{equation}
where $n>5$ is some positive integer. In this case, the linearised Hamiltonian constraint \eqref{eqn:Ham_linear} gets modified such that $q({\bf x})$ now has the ``right sign'', i.e. 
\begin{equation}
    q({\bf x})\equiv  \frac{1}{12}\psi_0^4\left( 5K^2 - 24(5-n)\pi \tilde{\rho}\psi_0^{-n}\right) ~> 0
\end{equation}
and the maximum principle applies.  This strategy is useful when one can specify $\rho$ as a fundamental quantity which is independent of $\psi_0$.  In this case, $\tilde{\rho}$ becomes the freely specifiable quantity and once the solution for $\psi$ is found, one reconstructs the physical energy density $\rho$ from \eqsref{eqn:rho_rescale}. As a consequence, a constant choice of $\tilde\rho$ generally results in an inhomogeneous profile for $\rho$. While the loss of one's ability to specify a particular configuration for $\rho$ can be an inconvenience at best, one can still take comfort that the rescaling is one-to-one and unambiguous.

However, when dealing with a fundamental field $\phi$, for example, the energy density $\rho$ itself is a function of $\phi$, $\Pi$, $V(\phi)$ and $\psi_0$. 
This implies that for any given $\tilde{\rho}$, there exist multiple possible configurations for $\phi$ and $\Pi$ which can reproduce the resulting (and generally spatially varying) $\rho$. If we add physical restrictions on the fields or their boundary conditions, the reconstruction of the field configuration may be difficult or in some cases, impossible. For example, if we start with a system with only potential energy density $\rho=V(\phi)$, $\partial_i \phi=0$, $\Pi=0$, rescaling either the energy density $\rho$ or equivalently the field $\phi$ by the conformal factor and imposing the value of the rescaled quantity $\tilde \phi$ will usually result in a space dependent configuration $\phi(x)\sim \tilde{\phi}\psi(x)^{-n}$. This will introduce additional gradient energy density into the system $\rho \sim \psi^{-4} (\partial_i \phi)^2 + V(\phi)$, which we may not want, and moreover may not be compatible with the boundary conditions (if, for example, we require $\phi=$ constant on the boundary).
\\

\section{\cttk: A new method to solve the ADM constraints}
\label{sec-ourmethod}

In this section we describe in detail our method for solving the constraints. 
The method largely follows the CTT approach -- for completeness we review the key elements in \ref{sec-method}. In the following we restrict ourselves to the case of conformally flat spacetimes ($\bar{\gamma}_{ij} = \delta_{ij}$), although we expect our method to work in the more general case. 

In the \cttk{} approach, we begin with the standard set of constraint equations as in the CTT method, but instead of solving for $\psi$ for a given $K$, we do the following:
\begin{itemize}
    \item In the Hamiltonian constraint we specify a profile for $\psi$ and solve for $K$.
    \item In the momentum constraint the contribution of $\partial_i K$ is then always non-zero and must be included as a source.
\end{itemize}
With this choice, the Hamiltonian constraint reduces from a 2nd order elliptic differential equation for $\psi$ to an \emph{algebraic} equation for $K$ where
\begin{empheq}[box=\fbox]{align}\label{eq:K_method1}
K^2 = 12\psi_0^{-5}\parpar\psi_0 +
\frac{3}{2}\psi_0^{-12} \bar{A}_{ij}\bar{A}^{ij} +24\pi\rho
\,,
\end{empheq}
with the freedom to choose between an overall collapsing or expanding $K=\pm \sqrt{K^2}$ spacetime\footnote{Provided that $K^2$ does not go to zero across a closed surface, $K$ should not change sign.}.  The advantage of this approach is apparent -- an algebraic equation is much simpler to solve than an elliptic different equation, evading the existence and uniqueness issues we discussed in Sec. \ref{sec-background}. Up to a sign change, a solution can always be found for $K$ as long as the RHS of  \eqref{eq:K_method1} is positive semi-definite.

Intuitively, such a choice determines the matter densities measured by normal observers on the initial hyperslices, which depend on the physical problem under consideration. Two canonical cases are cosmological spacetimes, where $\psi_0$ is simply a spatially constant value that determines the initial scale factor, and black hole spacetimes, where Brill-Lindquist type initial data of the form $\psi_0=\psibh=1+M/2r$ is used (\cite{Brill:1963yv}, see \ref{sec-bhs}). Note that in both these cases the Laplacian term in \eqref{eq:K_method1} is zero, which ensures that the RHS of \eqref{eq:K_method1} is never negative and hence a solutions always exists for $K$.

Since $K$ is spatially varying in general, then even in the case of zero momentum densities $S_i=0$, the momentum constraints will have a non-zero source, so they always need to be solved. The momentum constraint takes the usual linear form
\begin{empheq}[box=\fbox]{align}\label{eq:Mom}
\left(\bar{\Delta}_\mathrm{L}W\right)_i = \frac{2}{3}\psi_0^6\partial_i K + 8\pi\psi_0^{6} S_i
\,.
\end{empheq}
In general solving the momentum constraints in this form is straightforward, and $W_i$ can be obtained via the vector Laplacian decompositions as described in \ref{appendix:vec_lapl}.

The choice of initial guess for $W_i$ determines the solution that is picked out, which is unaffected by the addition of a constant to $W_i$. The same is not true in the case where different constant guesses for $\psi$ are chosen in the CTT approach, since they may change the physical solutions if the energy density is a function of $\psi$. Another advantage in the case of periodic spacetimes is that the integrability condition of the Hamiltonian constraint $\mathcal{I_H}$ introduced in the previous section is now trivially satisfied, whilst the equivalent condition for the momentum constraints $\mathcal{I}_{\mathcal{M}_i}$ only requires $K$ to be periodic -- satisfied by definition -- and gives similar restrictions on the choices of $S_i$ as in the CTT method. Note that the CTTK method, if the integrability conditions are initially satisfied, they will remain satisfied throughout all iteration steps of the solver, as $K$ will remain periodic (if the sources are periodic) and the second term in \eqref{eq:S_integrability} is unchanged throughout the solver iterations. Finally, we note that since for each iterative step we are solving a pair of algebraic  (\eqref{eq:K_method1}) and effectively a Poisson-like equation with sources (\eqref{eq:Mom}), the iterative solution is unique modulo the sign of $K$.

Operationally, the approach is as follows: We choose $\psi_0$ and an initial guess for $(W_i)_0$. We solve the Hamiltonian constraint using \eqref{eq:K_method1} (which now includes ($(W_i)_0$) and $\psi_0$ as sources) to find a value for $K_0$, which adds a source to the the momentum constraints (\eqref{eq:Mom}). This is now solved (using the methods of \ref{appendix:vec_lapl}) for $(W_i)_1$. We repeat this process starting now from ($\psi_0,~(W_i)_1$) and iterate to obtain $K_n$ and $(W_i)_n$ until the solutions converge\footnote{One could be concerned that the fact that, due to these iterations, the source in the momentum constraint contains contributions from second derivatives of $W_i$ via $\partial_i K$ which should be included in the principle part, but in practise we do not find that this gives any issues with convergence of the solutions.}, which they do rapidly, even from a naive initial guess of $W_i=0$, as demonstrated in Sec. \ref{sec-results}.

The idea of \cttk{}, setting $K$ algebraically, can be combined with the standard CTT method to produce a ``hybrid \cttk{}'' formalism that we have found yields an improvement in convergence rate (when compared to the ``pure'' \cttk{}) where there are contributions to the value of $\bar{A}_{ij}$ from non-matter sources (e.g. with Bowen-York data or other non-zero transverse-traceless components). This hybrid form is specified by the following choice:
\begin{empheq}[box=\fbox]{gather}
K^2 = 24\pi\rho \label{eq:Ham_BH1}
\,, \\
\partial^j\partial_j \psi=-\frac{1}{8}\psi^{-7} 
\bar{A}_{ij} \bar{A}^{ij}
\,, \label{eq:Ham_BH2}
\end{empheq}
where we have explicitly set the expansion $K$ equal to the energy density, while retaining the elliptic equation of $\psi$ sourced by the $\bar{A}_{ij}$ terms. The Hamiltonian constraint in this approach, \eqref{eq:Ham_BH2}, features with the ``right sign'' and thus always has unique solutions in the intermediate steps as discussed in detail in Sec. \ref{sec-background}, and the momentum constraints can be solved using the same techniques, \ref{appendix:vec_lapl}.

The disadvantage of using this hybrid approach is that the proper distance on the initial hyperslice is now not a fixed choice since the final value of $\psi \neq \psi_0$. As a result, if the sources $\rho$ and/or $S_i$ depends on $\psi$, e.g. in the case of a fundamental scalar field, one needs to recalculate $\rho$ and $S_i$ after obtaining $\psi$, and iterate until convergence is achieved. Depending on the physical problem of interest, this can be an acceptable compromise, particularly when the initial trial $\psi_0$ is close to the true solution. 

\section{Numerical validation}
\label{sec-results}

In this section we discuss the results after implementing the \cttk{} method using a multigrid approach built with the \textsc{chombo} library \cite{Chombo}.  We discuss two illustrative physical problems, covering the two key cases of periodic boundaries and asymptotically flat black hole spacetimes.  In both cases we use non-trivial density and momentum profiles derived from a real scalar field $\phi({\bf x})$ and conjugate momentum $\Pi({\bf x})$, so that
\begin{align}
\rho({\bf x}) &= \frac{1}{2}\psi^{-4}(\partial_i \phi)^2 + \frac{1}{2}\Pi^2 + V(\phi)
\,, \label{eq:rho_scalar}\\
S_i({\bf x}) &= -\Pi ~\partial_i\phi \label{eq:Si_scalar}
\,.
\end{align}

We compute the Hamiltonian ($\mathcal{H}$) and momentum ($\mathcal{M}\equiv\sqrt{\mathcal{M}_i^2}$) constraint errors via Eqns. \eqsref{eq:Ham_eq1} and \eqsref{eq:Mom_eq1} and iterate the initial condition solver until their global error norms have been appropriately reduced. We then perform convergence tests by comparing the local value of the constraints across the grid for low and high resolutions with $N_\mathrm{LR}$ and  $N_\mathrm{HR}$ number of grid points respectively. Both the Hamiltonian and momentum constraints must decrease with resolution as
\begin{equation}
    \lim_{\Delta x\rightarrow 0} \frac{\mathcal{H}_\mathrm{HR}}{\mathcal{H}_\mathrm{LR}} = \lim_{\Delta x\rightarrow 0} \frac{\mathcal{M}_\mathrm{HR}}{\mathcal{M}_\mathrm{LR}}  \approx \left(\frac{N_\mathrm{LR}}{N_\mathrm{HR}}\right)^n ,
\end{equation}
where $n$ is the order of convergence, limited by the stencils used\footnote{In our case $n=2$ due to the red-black linear algorithm in the solver.}.

\begin{figure}[t]
    \includegraphics[width=\columnwidth]{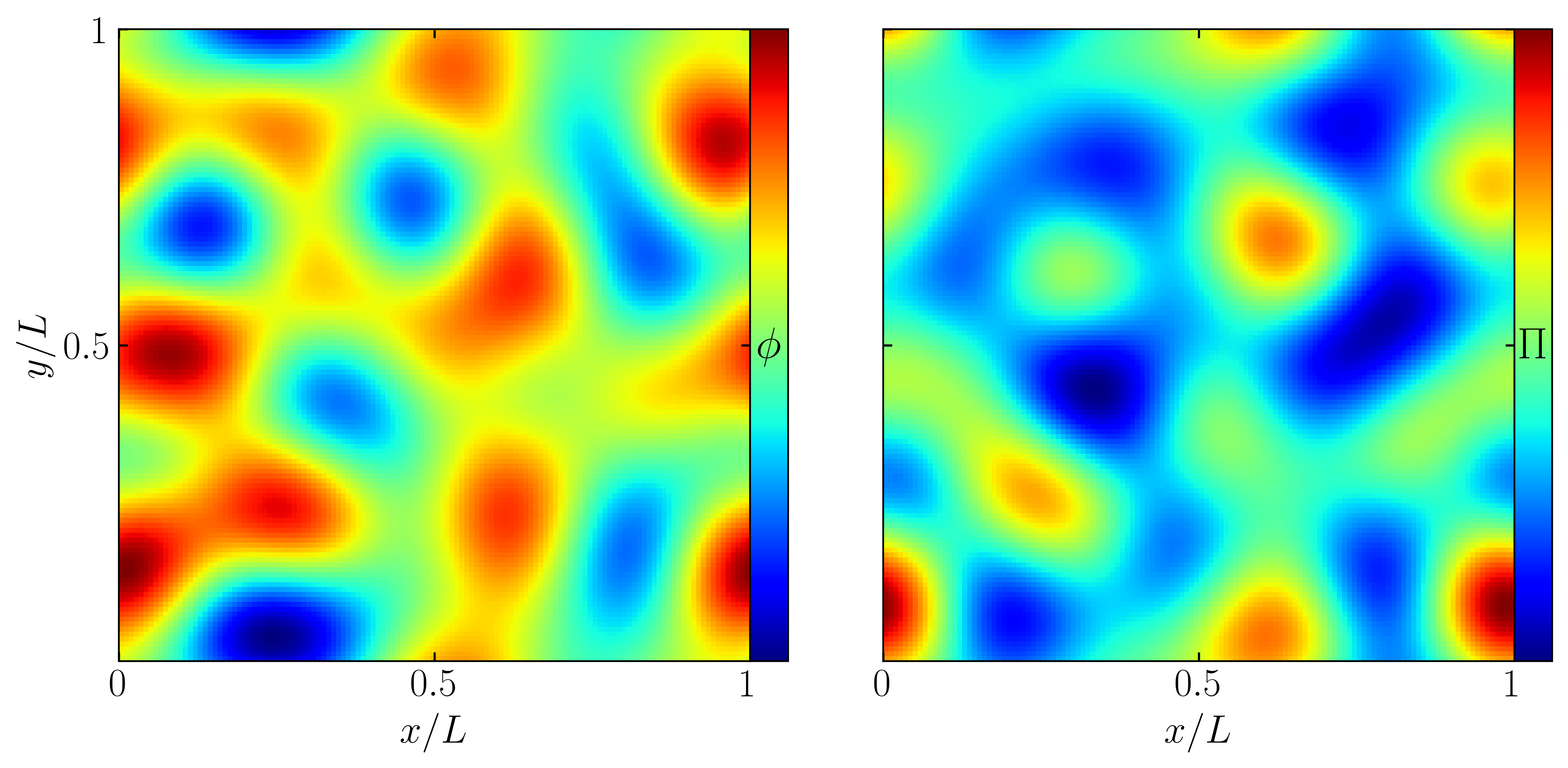}
    \caption{Gaussian random fields in $\phi$ and $\Pi$ in a periodic box of size $L=0.12\bar{H}^{-1}$ where $\bar{H}^{-1}$ is the Hubble length averaged over the initial hyperslice. The mass of the potential is $m = 3.45\bar{H}$ and the parameters of the perturbations are $(\phi_\mathrm{max},\phi_\mathrm{min}) = (-9.96,-10.03)\times 10^{-2}\mpl$ and $(\Pi_\mathrm{max},\Pi_\mathrm{min}) = (0.41,0.28)\mpl\bar{H}$. The colour scale on the plots is linear. The non-trivial profiles source both energy density $\rho$ and momentum density $S_i$ components of the stress-energy tensor.}
    \label{fig:panel_periodic}
\end{figure}

\subsection{Gaussian random scalar field and momentum with periodic boundary conditions}

We test the \cttk{} method in a box of length $L$ with periodic boundary conditions by simulating a set of Gaussian random fields in $\phi$ and $\Pi$ with a quadratic scalar potential $V(\phi) = m^2\phi^2/2$, as shown in Fig. \ref{fig:panel_periodic}. This field profile sources non-trivial stress-energy components $\rho$ and $S_i$, and we choose the field parameters so that modes propagate in both directions roughly equally in the spatial slice -- ensuring that the periodic integrability condition for $S_i$ is approximately satisfied.

We choose the conformal factor $\psi_0=1$ and thus fix the initial (and final) matter energy density. We start with the guess $W_i=0$, so that in the first iteration $\bar{A}_{ij}=0$. We then solve the Hamiltonian constraint for $K^2$ via \eqref{eq:K_method1}. This space dependent $K$, together with the non-zero $S_i$, now source the momentum constraint, which we solve by decomposing the vector Laplacian using the \emph{ansatz} for non-compact sources (Eqns. \eqsref{eq:vect_lapl_1} and  \eqref{eq:vect_lapl_2}). As noted above, this gives rise to linear Poisson equations for $W_i$ which admit an infinite number of solutions $W_i+\mathrm{const}$ when solved with periodic boundary conditions. Whilst these are physically equivalent since they give rise to the same $\bar{A}_{ij}\sim \partial_i W_j$, the solver will not converge if the freedom remains unchecked. We therefore pick out the closest to our initial guess by modifying the Laplace operator to $[\partial^j\partial_j]\rightarrow [\partial^j\partial_j  -c]$, where $c$ is a positive constant that we choose to be of order the solver tolerance \cite{Isenberg_1995,Curtis:2005va}\footnote{\ We are grateful to David Garfinkle for pointing out this trick to us (in the context of solving the non-linear Hamiltonian constraint for periodic spacetimes), which he in turn credited to Jim Isenberg.}.
We then reconstruct $\bar{A}_{ij}$ from the solution of $W_i$, which provides an additional source for $K^2$ in the next iteration of the Hamiltonian constraint. The process is iterated until a set tolerance on the error is achieved.

\begin{figure}[t]
    \includegraphics[width=\columnwidth]{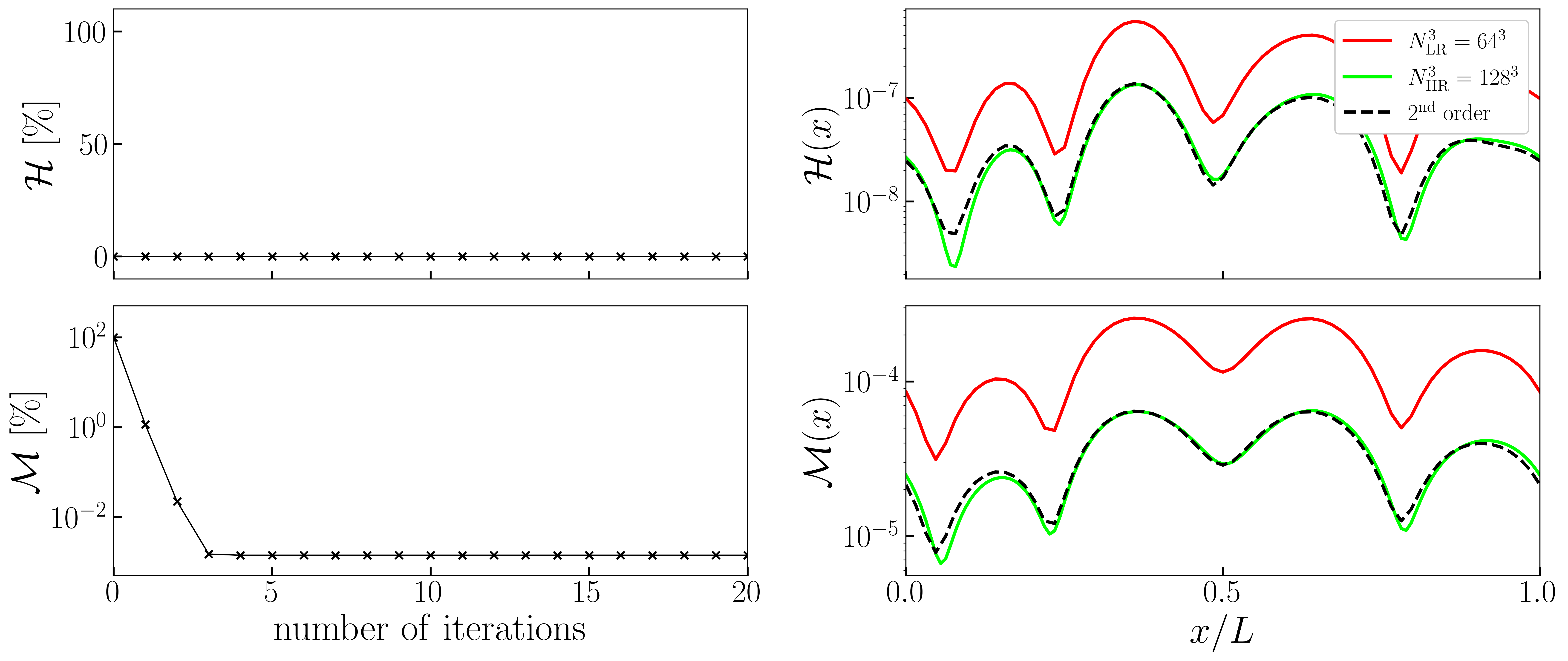}
    \caption{Convergence test for the initial Gaussian random fields in a box of size $L$ with periodic boundary conditions. The left panels show the normalised global convergence of the Hamiltonian and momentum constraints norms with solver iterations. Note that in this approach, the Hamiltonian constraint is algebraically solved and so it remains satisfied at every iteration. The right panels show the final local convergence, the Hamiltonian and momentum constraint errors across the grid $(x,L/2,L/2)$ for two resolutions, with very good agreement to the expected 2nd order.}
    \label{fig:constraints_periodic}
\end{figure}

In the left panels of Fig. \ref{fig:constraints_periodic} we plot the evolution of the Hamiltonian and momentum constraint error norms with the number of iterations. Note that as we are solving the Hamiltonian constraint algebraically, the error is always zero (to numerical precision) by construction. The momentum constraint converges within a few iterations. In the right panels we plot the final local constraint errors for low and high resolutions, showing agreement with the expected 2nd order convergence. Note that even though the solver reduces the global error and saturates within a few of iterations, we run it for longer so that local constraint errors that do not dominate are converged too.

\subsection{Rotating scalar cloud around a spinning black hole with asymptotically flat boundary conditions}

\begin{figure}[t]
    \includegraphics[width=\columnwidth]{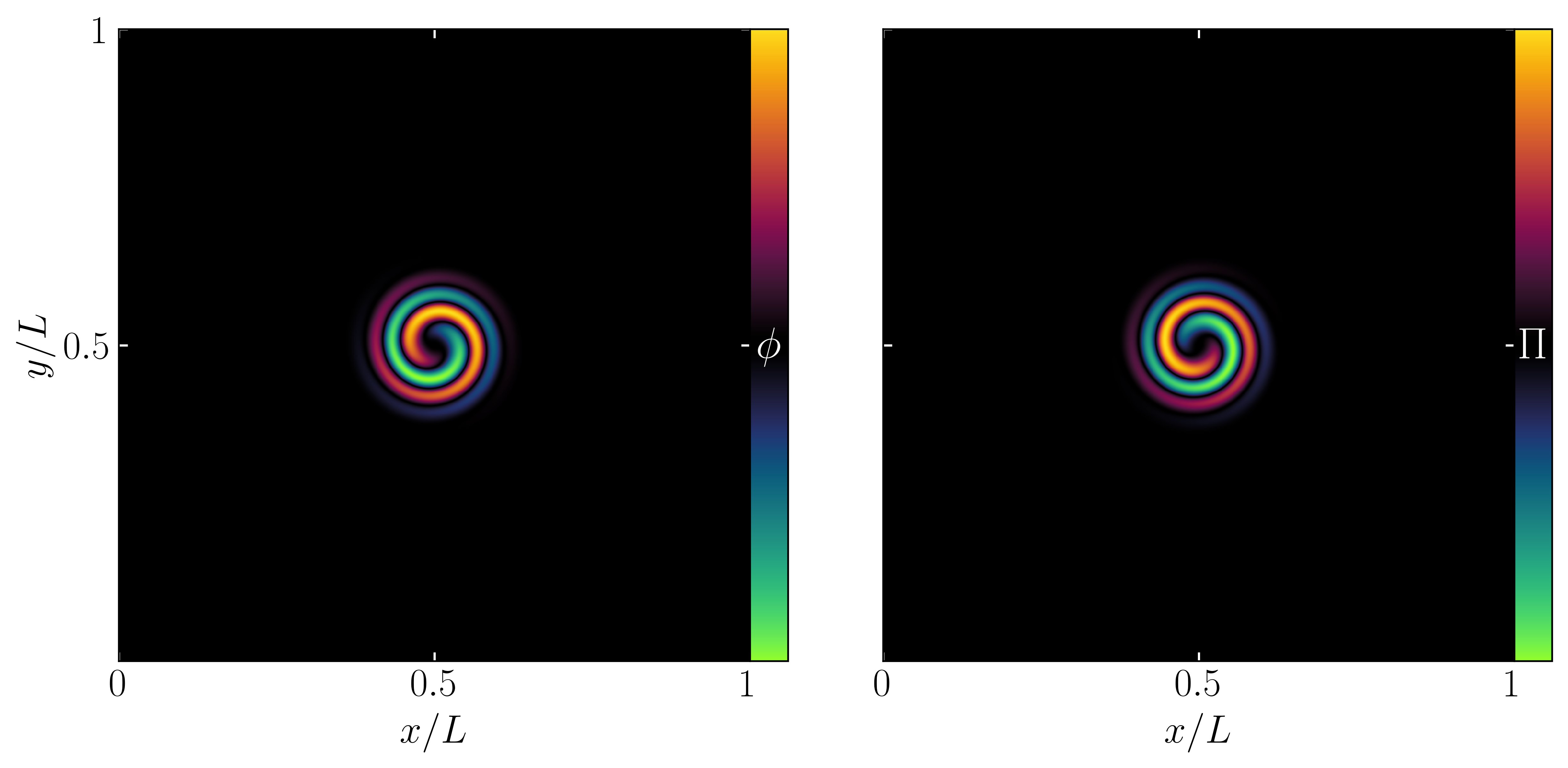}
    \caption{Field profiles for the scalar field $\phi$ and momentum $\Pi$ mimicking a rotating scalar cloud around a spinning black hole of mass $M$ and dimensionless spin parameter $a/M=0.5$, in a box of size $L = 128 M$. The mass of the potential is $m = 10^{-2} M^{-1}$ and the parameters of the cloud are $\Delta\phi = 10^{-4}\mpl$, $\Delta\Pi = 10^{-5} \mpl M^{-1}$ and $\sigma = 5M^{-1}$. The colour scale on the plots are linear.  These profiles induce non-trivial energy and momentum density components of the stress-energy tensor.}
    \label{fig:panel_fixed}
\end{figure}

We also test the hybrid \cttk{} method (i.e. Eqns. \eqsref{eq:Ham_BH1} and \eqsref{eq:Ham_BH2}) and our solver for the case of a scalar cloud with potential $V(\phi) = m^2\phi^2/2$ that is rotating around a spinning black hole. Scalar fields are popular models to describe the growth of a wave dark matter cloud around a black hole \cite{Bamber:2022pbs}, and are also relevant for superradiant instabilities \cite{Brito:2015oca}. In Fig. \ref{fig:panel_fixed} we plot the field profiles that we use to mimic the rotating cloud and are given by
\begin{align}
    \phi({\bf x}) &= \Delta\phi ~ r^2\exp\left[-r^2/\sigma^2\right] \sin(\alpha \cos(\beta \sqrt{x^2+y^2}) - \kappa \varphi) \\
    \Pi({\bf x}) &= \Delta\Pi ~ r^2\exp\left[-r^2/\sigma^2\right] \cos(\alpha \cos(\beta \sqrt{x^2+y^2}) - \kappa \varphi)
\end{align}
where $r^2=\sqrt{x^2+y^2+z^2}$ is the spherical radial coordinate and $\varphi$ is the azimuthal angle. The set of parameters ($\alpha,~\beta,~\kappa,~\sigma$) determine the precise shape of the cloud. 

We use the hybrid \cttk{} approach, where we choose $K^2$ to absorb only the matter terms, and linearise \eqref{eq:Ham_BH2} for the conformal factor $u_{n+1} = \psi_{n+1}- \psi_{n}$ solving it iteratively
\begin{equation} 
    \left(\parpar  -\frac{7}{8}\psi_n^{-8} 
    \bar{A}_{ij}\bar{A}^{ij}\right) u_{n+1}=-\frac{1}{8}\psi_n^{-7}  
    \bar{A}_{ij}\bar{A}^{ij} - \parpar\psi_n \label{eq:linearised_Ham_BH2}
\,,
\end{equation}
We expand $\bar{A}^{ij}=\bar{A}^{ij}_\mathrm{BH} + A^{ij}_*$ (and $W^i = W^i_\mathrm{BH} + W^i_*$), choose the initial guess $\psi_\mathrm{ini}=\psibh$ and $\bar{A}^{ij}_\mathrm{BH}$ to be the Bowen-York solution for a spinning black hole with mass $M$ (\eqref{eq:bowenyork}), which satisfies $\parpar \psibh = \partial_j \bar{A}^{ij}_\mathrm{BH}=0$.
The quantities $u$ and $\bar{A}^{ij}_*$ are the corrections we solve for. We start by solving the momentum constraints sourced by the spatially varying $K$, choosing $W^i_*=0$ as a starting guess for the correction to the Bowen-York data (note that the total $W^i\neq 0$, but we can ignore the Bowen-York $W^i_\mathrm{BH}$ in the momentum constraints since they are linear and it is a solution) and using the compact source \emph{ansatz} \eqref{eq:compact_ansatz}. We then reconstruct the corrected $\bar{A}^{ij}$ and solve the Hamiltonian constraint for $u$. After we correct our trial solution $\psi_{n+1} \rightarrow \psi_n + u_{n+1}$, we recompute the new $K^2$ and iterate the same process until we converge to the final solution when the right hand side of \eqref{eq:linearised_Ham_BH2} achieves a set tolerance. As in the previous subsection, we run the solver for a few more iterations after the global error saturates, to make sure that the local constraints are converged too.

\begin{figure}[t]
    \includegraphics[width=\columnwidth]{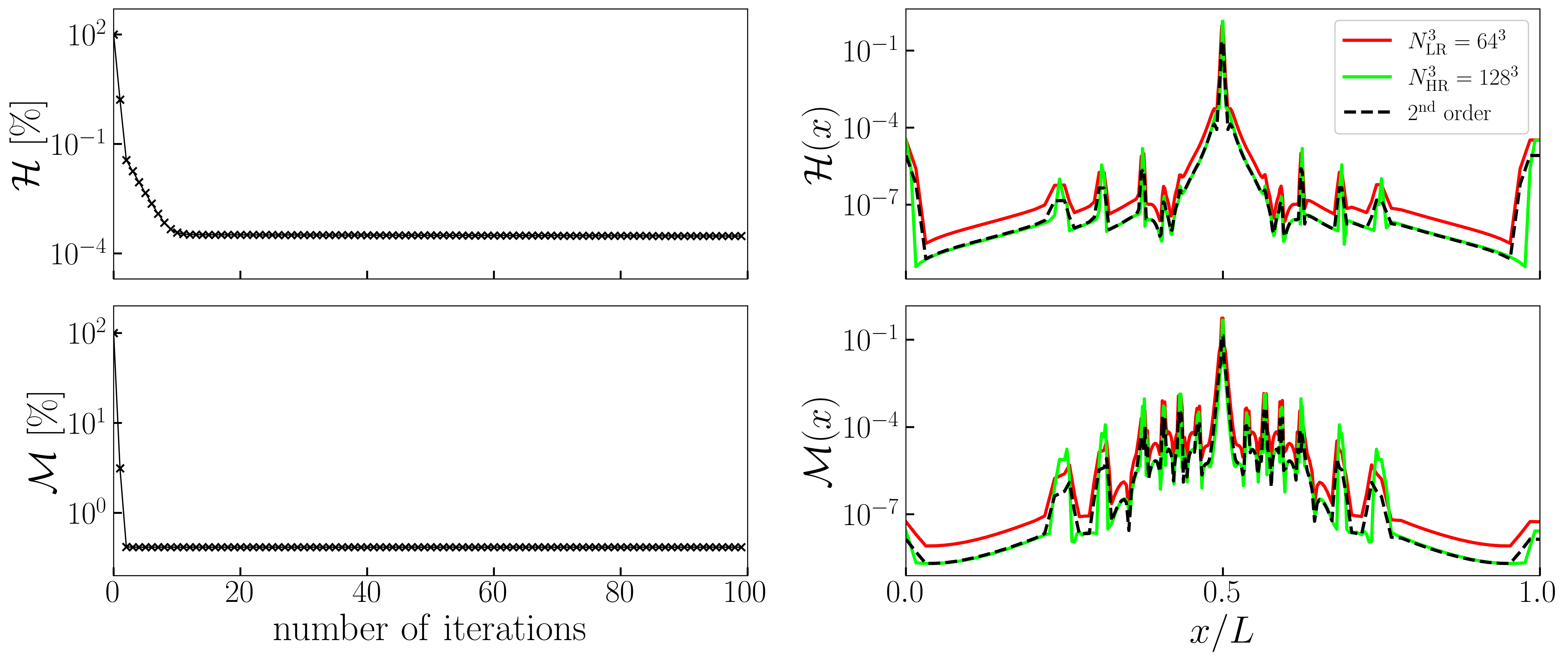}
    \caption{Convergence test for a rotating cloud around a spinning black hole. The left panels show the global convergence of the norm of the Hamiltonian and momentum constraints. The right panels show the final local convergence of the Hamiltonian and momentum constraints across the grid, in perfect agreement with the expected 2nd order. The sharp spikes correspond to the location of the refinement boundaries.}
    \label{fig:constraints_fixedBH}
\end{figure}

We impose asymptotically flat boundary conditions by using extrapolating boundary conditions for $u$ and $W_i$, so that the asymptotic solution is consistent with having zero source terms to the Laplacian near the boundary. We choose three refinement levels, and enforce the condition that regridding boundaries of the low and high resolution runs approximately coincide. This is important in simulations with Adaptive Mesh Refinement, since convergence tests aim to compare the error in the same region for different resolutions, and if the regridding boundaries do not coincide, the resolution in some regions might not have increased consistently. Fig. \ref{fig:constraints_fixedBH} confirms the expected 2nd order convergence.

\section{Conclusions}
\label{sec-conclude}

In this paper we proposed a new method for solving the ADM constraints based on the CTT method. In this \cttk{} approach, rather than choosing constant mean curvature slices, one chooses the conformal factor and solves for a spatially varying $K$. This has the advantage of avoiding potential issues with non-uniqueness and existence of solutions in the Hamiltonian constraint while keeping control of the specified initial data sources. The price to pay is that the Hamiltonian and momentum constraints become unavoidably coupled and must be simultaneously solved even in cases with zero momentum density sources, but this does not present any significant technical challenge. The method is very stable -- we have found that it gives rapid convergence for a number of physical problems of interest, including mixed systems of black holes in scalar field environments and strongly inhomogeneous closed spacetimes. 

It can therefore be a useful alternative to existing methods, in particular in the case of spacetimes containing fundamental fields. For example, it may help stabilise the initial conditions for superposed boson stars, where one would like to retain the ``unperturbed'' conformal factor profiles for individual stars \cite{Helfer:2018vtq,Helfer:2021brt}. It may also offer advantages in more traditional astrophysical scenarios including vacuum spacetimes and perfect fluids, where it would keep the mass of the system fixed with corrections to their spin and/or momentum when solving the constraints\footnote{~ We are grateful to Mark Hannam for highlighting this advantage.}.

Whilst we implemented our approach as a variation of the CTT method, we expect that it should work equally well as an amendment to XCTS,
and we plan to expand the method to this formalism in the future, along with the ability to choose a non-flat conformal metric. We also plan to make our implementation publicly available as part of the \grchombo{} code \cite{Clough:2015sqa,Andrade:2021rbd}.

%%%%%%%%%%%%%%%%%%%%%%%%%%%%%%%%%%%%%%%%%%%%%%%%%%%%%%%%%%%%%%%%%%%%%%%%%%%%%%%%%%%%%%%
\section*{Acknowledgments}
%%%%%%%%%%%%%%%%%%%%%%%%%%%%%%%%%%%%%%%%%%%%%%%%%%%%%%%%%%%%%%%%%%%%%%%%%%%%%%%%%%%%%%%
We acknowledge useful conversations with David Garfinkle and Mark Hannam. We would also like to thank Francesco Muia for the Gaussian scalar field profiles. This project has received funding from the European Research Council (ERC) under
the European Union’s Horizon 2020 research and innovation programme (grant agreement No 693024). JCA acknowledges funding from the Beecroft Trust and The Queen’s College via an extraordinary Junior Research Fellowship (eJRF). KC acknowledges funding from the ERC, and an STFC Ernest Rutherford Fellowship project reference ST/V003240/1.

This  work  was  performed  using  the Leibnitz Supercomputing Centre SuperMUC-NG under  PRACE grant Tier-0 Proposal 2018194669, on the J\"ulich Supercomputing Center JUWELS HPC under PRACE grant Tier-0 Proposal 2020225359, COSMA7 in Durham and Leicester DiAL HPC under DiRAC RAC13 Grant ACTP238, and the Cambridge Data Driven CSD3 facility which is operated by the University of Cambridge Research Computing on behalf of the STFC DiRAC HPC Facility.  The  DiRAC  component  of CSD3 was funded by BEIS capital funding via STFC capital grants ST/P002307/1 and ST/R002452/1 and STFC operations grant ST/R00689X/1.

\bibliographystyle{iopart-num}
\bibliography{mybib}

\providecommand{\newblock}{}
\begin{thebibliography}{10}
\expandafter\ifx\csname url\endcsname\relax
  \def\url#1{{\tt #1}}\fi
\expandafter\ifx\csname urlprefix\endcsname\relax\def\urlprefix{URL }\fi
\providecommand{\eprint}[2][]{\url{#2}}

\bibitem{Arnowitt:1962hi}
Arnowitt R~L, Deser S and Misner C~W 2008 {\em Gen. Rel. Grav.\/} {\bf 40}
  1997--2027 (\textit{Preprint} \eprint{gr-qc/0405109})

\bibitem{Mertens:2015ttp}
Mertens J~B, Giblin J~T and Starkman G~D 2016 {\em Phys. Rev. D\/} {\bf 93}
  124059 (\textit{Preprint} \eprint{1511.01106})

\bibitem{York:1998hy}
York Jr J~W 1999 {\em Phys. Rev. Lett.\/} {\bf 82} 1350--1353
  (\textit{Preprint} \eprint{gr-qc/9810051})

\bibitem{Pfeiffer:2002iy}
Pfeiffer H~P and York Jr J~W 2003 {\em Phys. Rev. D\/} {\bf 67} 044022
  (\textit{Preprint} \eprint{gr-qc/0207095})

\bibitem{Baumgarte:2010ndz}
Baumgarte T~W and Shapiro S~L 2010 {\em {Numerical Relativity: Solving
  Einstein's Equations on the Computer}\/} (Cambridge University Press)

\bibitem{Gourgoulhon:2007ue}
Gourgoulhon E 2007  (\textit{Preprint} \eprint{gr-qc/0703035})

\bibitem{Alcubierre:2008co}
Alcubierre M 2008 {\em {Introduction to 3+1 Numerical Relativity}\/}
  International Series of Monographs on Physics (Oxford: Oxford University
  Press) ISBN 9780199205677

\bibitem{Tichy:2016vmv}
Tichy W 2017 {\em Rept. Prog. Phys.\/} {\bf 80} 026901 (\textit{Preprint}
  \eprint{1610.03805})

\bibitem{East:2012zn}
East W~E, Ramazanoglu F~M and Pretorius F 2012 {\em Phys. Rev. D\/} {\bf 86}
  104053 (\textit{Preprint} \eprint{1208.3473})

\bibitem{Assumpcao:2021fhq}
Assumpcao T, Werneck L~R, Jacques T~P and Etienne Z~B 2022 {\em Phys. Rev. D\/}
  {\bf 105} 104037 (\textit{Preprint} \eprint{2111.02424})

\bibitem{OMurchadha:1973byk}
O'Murchadha N and York Jr J~W 1973 {\em J. Math. Phys.\/} {\bf 14} 1551--1557

\bibitem{Pfeiffer:2005jf}
Pfeiffer H~P and York Jr J~W 2005 {\em Phys. Rev. Lett.\/} {\bf 95} 091101
  (\textit{Preprint} \eprint{gr-qc/0504142})

\bibitem{Baumgarte:2006ug}
Baumgarte T~W, Murchadha N~O and Pfeiffer H~P 2007 {\em Phys. Rev. D\/} {\bf
  75} 044009 (\textit{Preprint} \eprint{gr-qc/0610120})

\bibitem{Walsh:2006au}
Walsh D~M 2007 {\em Class. Quant. Grav.\/} {\bf 24} 1911--1926
  (\textit{Preprint} \eprint{gr-qc/0610129})

\bibitem{Cordero-Carrion:2008grk}
Cordero-Carrion I, Cerda-Duran P, Dimmelmeier H, Jaramillo J~L, Novak J and
  Gourgoulhon E 2009 {\em Phys. Rev. D\/} {\bf 79} 024017 (\textit{Preprint}
  \eprint{0809.2325})

\bibitem{Garfinkle:2020iup}
Garfinkle D and Mead L 2020 {\em Phys. Rev. D\/} {\bf 102} 044022
  (\textit{Preprint} \eprint{2006.16360})

\bibitem{York:1978gql}
York Jr J~W 1978 {Kinematics and Dynamics of General Relativity} {\em {Workshop
  on Sources of Gravitational Radiation}\/}

\bibitem{Suh:2016ctx}
Suh I, Mathews G~J, Haywood J~R and Lan N~Q 2017 {\em Adv. Astron.\/} {\bf
  2017} 6127031 (\textit{Preprint} \eprint{1601.01460})

\bibitem{1979Smarr}
{Smarr} L~L, {Epstein} R and {Clark} J~P~A 1979 {\em {Sources of gravitational
  radiation. Proceedings of the Battelle Seattle workshop, July 24 - August 4,
  1978.}\/}

\bibitem{Bentivegna:2013xna}
Bentivegna E 2014 {\em Class. Quant. Grav.\/} {\bf 31} 035004
  (\textit{Preprint} \eprint{1305.5576})

\bibitem{East:2015ggf}
East W~E, Kleban M, Linde A and Senatore L 2016 {\em JCAP\/} {\bf 09} 010
  (\textit{Preprint} \eprint{1511.05143})

\bibitem{Clough:2016ymm}
Clough K, Lim E~A, DiNunno B~S, Fischler W, Flauger R and Paban S 2017 {\em
  JCAP\/} {\bf 09} 025 (\textit{Preprint} \eprint{1608.04408})

\bibitem{Clough:2017efm}
Clough K, Flauger R and Lim E~A 2018 {\em JCAP\/} {\bf 05} 065
  (\textit{Preprint} \eprint{1712.07352})

\bibitem{Aurrekoetxea:2019fhr}
Aurrekoetxea J~C, Clough K, Flauger R and Lim E~A 2020 {\em JCAP\/} {\bf 05}
  030 (\textit{Preprint} \eprint{1910.12547})

\bibitem{Joana:2020rxm}
Joana C and Clesse S 2021 {\em Phys. Rev. D\/} {\bf 103} 083501
  (\textit{Preprint} \eprint{2011.12190})

\bibitem{Yoo:2018pda}
Yoo C~M, Ikeda T and Okawa H 2019 {\em Class. Quant. Grav.\/} {\bf 36} 075004
  (\textit{Preprint} \eprint{1811.00762})

\bibitem{Yoo:2020lmg}
Yoo C~M, Harada T and Okawa H 2020 {\em Phys. Rev. D\/} {\bf 102} 043526
  (\textit{Preprint} \eprint{2004.01042})

\bibitem{deJong:2021bbo}
de~Jong E, Aurrekoetxea J~C and Lim E~A 2022 {\em JCAP\/} {\bf 03} 029
  (\textit{Preprint} \eprint{2109.04896})

\bibitem{Garfinkle:2008ei}
Garfinkle D, Lim W~C, Pretorius F and Steinhardt P~J 2008 {\em Phys. Rev. D\/}
  {\bf 78} 083537 (\textit{Preprint} \eprint{0808.0542})

\bibitem{Cook:2020oaj}
Cook W~G, Glushchenko I~A, Ijjas A, Pretorius F and Steinhardt P~J 2020 {\em
  Phys. Lett. B\/} {\bf 808} 135690 (\textit{Preprint} \eprint{2006.01172})

\bibitem{Ijjas:2020dws}
Ijjas A, Cook W~G, Pretorius F, Steinhardt P~J and Davies E~Y 2020 {\em JCAP\/}
  {\bf 08} 030 (\textit{Preprint} \eprint{2006.04999})

\bibitem{Ijjas:2021wml}
Ijjas A, Pretorius F, Steinhardt P~J and Sullivan A~P 2021 {\em Phys. Lett.
  B\/} {\bf 820} 136490 (\textit{Preprint} \eprint{2104.12293})

\bibitem{Ijjas:2021zyf}
Ijjas A, Pretorius F, Steinhardt P~J and Garfinkle D 2021 {\em JCAP\/} {\bf 12}
  030 (\textit{Preprint} \eprint{2109.09768})

\bibitem{Widdicombe:2018oeo}
Widdicombe J~Y, Helfer T, Marsh D~J~E and Lim E~A 2018 {\em JCAP\/} {\bf 10}
  005 (\textit{Preprint} \eprint{1806.09367})

\bibitem{Giblin:2019nuv}
Giblin J~T and Tishue A~J 2019 {\em Phys. Rev. D\/} {\bf 100} 063543
  (\textit{Preprint} \eprint{1907.10601})

\bibitem{Kou:2019bbc}
Kou X~X, Tian C and Zhou S~Y 2021 {\em Class. Quant. Grav.\/} {\bf 38} 045005
  (\textit{Preprint} \eprint{1912.09658})

\bibitem{Kou:2021bij}
Kou X~X, Mertens J~B, Tian C and Zhou S~Y 2022 {\em Phys. Rev. D\/} {\bf 105}
  123505 (\textit{Preprint} \eprint{2112.07626})

\bibitem{Joana:2022uwc}
Joana C 2022 {\em Phys. Rev. D\/} {\bf 106} 023504 (\textit{Preprint}
  \eprint{2202.07604})

\bibitem{Yoo:2012jz}
Yoo C~M, Abe H, Nakao K~i and Takamori Y 2012 {\em Phys. Rev. D\/} {\bf 86}
  044027 (\textit{Preprint} \eprint{1204.2411})

\bibitem{Yoo:2013yea}
Yoo C~M, Okawa H and Nakao K~i 2013 {\em Phys. Rev. Lett.\/} {\bf 111} 161102
  (\textit{Preprint} \eprint{1306.1389})

\bibitem{Yoo:2014boa}
Yoo C~M and Okawa H 2014 {\em Phys. Rev. D\/} {\bf 89} 123502
  (\textit{Preprint} \eprint{1404.1435})

\bibitem{Giblin:2015vwq}
Giblin J~T, Mertens J~B and Starkman G~D 2016 {\em Phys. Rev. Lett.\/} {\bf
  116} 251301 (\textit{Preprint} \eprint{1511.01105})

\bibitem{Bentivegna:2015flc}
Bentivegna E and Bruni M 2016 {\em Phys. Rev. Lett.\/} {\bf 116} 251302
  (\textit{Preprint} \eprint{1511.05124})

\bibitem{Bentivegna:2016fls}
Bentivegna E, Korzy\'nski M, Hinder I and Gerlicher D 2017 {\em JCAP\/} {\bf
  03} 014 (\textit{Preprint} \eprint{1611.09275})

\bibitem{Bentivegna:2018koh}
Bentivegna E, Clifton T, Durk J, Korzy\'nski M and Rosquist K 2018 {\em Class.
  Quant. Grav.\/} {\bf 35} 175004 (\textit{Preprint} \eprint{1801.01083})

\bibitem{Giblin:2019pql}
Giblin J~T, Mertens J~B, Starkman G~D and Tian C 2019 {\em Class. Quant.
  Grav.\/} {\bf 36} 195009 (\textit{Preprint} \eprint{1903.01490})

\bibitem{Corman:2022rqo}
Corman M, East W~E and Ripley J~L 2022  (\textit{Preprint} \eprint{2206.08466})

\bibitem{Brill:1963yv}
Brill D~R and Lindquist R~W 1963 {\em Phys. Rev.\/} {\bf 131} 471--476

\bibitem{Chombo}
Adams M, Colella P, Graves D~T, Johnson J~N, Keen N~D, Ligocki T~J, Martin D~F,
  McCorquodale P~W, Modiano D, Schwartz P~O, Sternberg T~D and Straalen B
  {Chombo Software Package for AMR Applications - Design Document, Lawrence
  Berkeley National Laboratory Technical Report LBNL-6616E.}

\bibitem{Isenberg_1995}
Isenberg J 1995 {\em Classical and Quantum Gravity\/} {\bf 12} 2249--2274
  \urlprefix\url{https://doi.org/10.1088/0264-9381/12/9/013}

\bibitem{Curtis:2005va}
Curtis J and Garfinkle D 2005 {\em Phys. Rev. D\/} {\bf 72} 064003
  (\textit{Preprint} \eprint{gr-qc/0506107})

\bibitem{Bamber:2022pbs}
Bamber J, Aurrekoetxea J~C, Clough K and Ferreira P~G 2022  (\textit{Preprint}
  \eprint{2210.09254})

\bibitem{Brito:2015oca}
Brito R, Cardoso V and Pani P 2015 {\em Lect. Notes Phys.\/} {\bf 906}
  pp.1--237 (\textit{Preprint} \eprint{1501.06570})

\bibitem{Helfer:2018vtq}
Helfer T, Lim E~A, Garcia M~A~G and Amin M~A 2019 {\em Phys. Rev. D\/} {\bf 99}
  044046 (\textit{Preprint} \eprint{1802.06733})

\bibitem{Helfer:2021brt}
Helfer T, Sperhake U, Croft R, Radia M, Ge B~X and Lim E~A 2022 {\em Class.
  Quant. Grav.\/} {\bf 39} 074001 (\textit{Preprint} \eprint{2108.11995})

\bibitem{Clough:2015sqa}
Clough K, Figueras P, Finkel H, Kunesch M, Lim E~A and Tunyasuvunakool S 2015
  {\em Class. Quant. Grav.\/} {\bf 32} 245011 (\textit{Preprint}
  \eprint{1503.03436})

\bibitem{Andrade:2021rbd}
Andrade T {\em et~al.\/} 2021 {\em J. Open Source Softw.\/} {\bf 6} 3703
  (\textit{Preprint} \eprint{2201.03458})

\bibitem{Bowen1979GeneralFF}
Bowen J 1979 {\em General Relativity and Gravitation\/} {\bf 11} 227--231

\bibitem{Bowen:1980yu}
Bowen J~M and York Jr J~W 1980 {\em Phys. Rev. D\/} {\bf 21} 2047--2056

\bibitem{Shibata:1999wi}
Shibata M 1999 {\em Prog. Theor. Phys.\/} {\bf 101} 1199--1233
  (\textit{Preprint} \eprint{gr-qc/9905058})

\bibitem{Nakamura:1998qd}
Nakamura T and Oohara K~i 1998 {A Way to 3-D numerical relativity: Coalescing
  binary neutron stars} {\em {Numerical Astrophysics 1998 (NAP 98)}\/}
  (\textit{Preprint} \eprint{gr-qc/9812054})

\bibitem{Brandt:1997tf}
Brandt S and Bruegmann B 1997 {\em Phys. Rev. Lett.\/} {\bf 78} 3606--3609
  (\textit{Preprint} \eprint{gr-qc/9703066})

\end{thebibliography}
\clearpage

\appendix

\section{The CTT decomposition}
\label{sec-method}

In this appendix we review the methodology of the Conformal Transverse-Traceless (CTT) decomposition, following closely the treatments in \cite{Gourgoulhon:2007ue, Alcubierre:2008co,Baumgarte:2010ndz}. 

\subsection{York-Lichnerowicz decomposition of the metric quantities}

To solve the constraint equations it is convenient to perform the York-Lichnerowicz conformal decomposition, in which one writes the spatial metric $\gamma_{ij}$ as a product of a conformal factor $\psi$ and a background metric $\bar{\gamma}_{ij}$,
\begin{equation}
\gamma_{ij} = \psi^4 \bar{\gamma}_{ij}
\,,
\end{equation}
with $\bar{\gamma}=\mathrm{det}\bar{\gamma}_{ij}=1$. The conformal factor absorbs the overall scale of the metric and leaves five degrees of freedom in the conformally related metric $\bar{\gamma}_{ij}$. One can show that given this choice, the Hamiltonian constraint reduces to
\begin{equation}
8\bar{D}^2\psi-\psi\bar{R}-\psi^5 K^2 + \psi^5 K_{ij}K^{ij} = -16\pi\psi^5\rho
\,,
\end{equation}
where $\bar{D}^2=\bar{\gamma}^{ij}\bar{D}_i\bar{D}_j$ is the covariant Laplace operator associated with $\bar{\gamma}_{ij}$. Given a choice of the conformal metric $\bar{\gamma}_{ij}$ and some specification of $K_{ij}$, the Hamiltonian constraint results in a Poisson equation that must be solved for the conformal factor $\psi$. 

One also separates the extrinsic curvature tensor into its trace $K$ and traceless part $A_{ij}$ as
\begin{equation}
K_{ij} = A_{ij} + \frac{1}{3}\gamma_{ij}K
\,,
\end{equation}
and further decomposes $A_{ij}$ as
\begin{align}
A_{ij} &= \psi^{-2}\bar{A}_{ij}
\,.
\end{align}
Using that $D_j A^{ij}=\psi^{-10}\bar{D}_j\bar{A}^{ij}$ the constraints become
\begin{align}
8\bar{D}^2\psi - \psi \bar{R} - \frac{2}{3}\psi^5 K^2 +\psi^{-7} \bar{A}_{ij}\bar{A}^{ij} = -16\pi\psi^5\rho, \\
\bar{D}_j \bar{A}^{ij} - \frac{2}{3}\psi^6 \bar{\gamma}^{ij}\bar{D}_j K = 8\pi\psi^{10} S^i
\,. \label{eqn:Ham_constraint}
\end{align}

Any symmetric, traceless tensor such as $\bar{A}^{ij}$ can be split into a transverse-traceless part that is divergence free and a longitudinal part that can be written in terms of the gradients of a vector. That is,
\begin{equation}
\bar{A}^{ij} = \bar{A}^{ij}_\mathrm{TT} + \bar{A}^{ij}_\mathrm{L}
\,,
\end{equation}
where the transverse part satisfies $\bar{D}_j \bar{A}^{ij}_\mathrm{TT}=0$ and the longitudinal part satisfies
\begin{equation}\label{eq:Aij_L}
\bar{A}^{ij}_\mathrm{L} = \bar{D}^i W^j + \bar{D}^j W^i - \frac{2}{3}\bar{\gamma}^{ij} \bar{D}_k W^k
\,.
\end{equation}
We can now write the divergence of $\bar{A}^{ij}$ as
\begin{equation}\label{eq:vec_lap_eq}
\bar{D}_j \bar{A}^{ij} = \bar{D}_j \bar{A}^{ij}_\mathrm{L}
= \bar{D}^2 W^i + \frac{1}{3}\bar{D}^i\left(\bar{D}_j W^j\right)
+ \bar{R}^i_{~j}W^j \equiv \left(\bar{\Delta}_\mathrm{L}W\right)^i
\,,
\end{equation}
where $\bar{\Delta}_\mathrm{L}$ is the vector Laplacian. Hence, the momentum constraints are rewritten as
\begin{equation}
\left(\bar{\Delta}_\mathrm{L}W\right)^i - \frac{2}{3}\psi^6\bar{\gamma}^{ij}\bar{D}_j K = 8\pi\psi^{10} S^i
\,,
\end{equation}
while in the Hamiltonian constraint the term in $\bar{A}_{ij}$ must be reconstructed from $W^i$ and $\bar{A}^{ij}_\mathrm{TT}$.

In the standard CTT approach, we choose the conformally related metric $\bar{\gamma}_{ij}$ to be flat, the mean curvature $K$ to be a spatially constant value. Given these choices, one solves the Hamiltonian and momentum constraints for $\psi$ and the vector potential $W^i$, and then reconstructs the physical solutions\footnote{We can check at this point the degrees of freedom: We started with $6$ in $\gamma_{ij}$ and $6$ in $K_{ij}$. Now we have $1$ in $\psi$, $5$ in $\bar{\gamma}_{ij}$, $1$ in $K$, $2$ in $\bar{A}^{ij}_\mathrm{TT}$ and $3$ in $\bar{A}^{ij}_\mathrm{L}$. So if we specify the $8$ degrees of freedom in $\bar{\gamma}_{ij}$, $K$ and $\bar{A}^{ij}_\mathrm{TT}$, the four constraint equations will fix the remaining $4$ degrees of freedom in $\psi$ and $\bar{A}^{ij}_\mathrm{L}$.} $\gamma_{ij}$ and $K_{ij}$.

\subsection{Solving the vector Laplacian}
\label{appendix:vec_lapl}

In this work we restrict ourselves to conformally flat spacetimes ($\bar{\gamma}_{ij} = \delta_{ij}$, $\bar{R}=0$) where the vector Laplacian $(\bar{\Delta}_\mathrm{L}W)^i$ in Cartesian coordinates reduces to
\begin{equation}\label{eq:vec_lap_eq_flat}
\left(\bar{\Delta}_\mathrm{L}W\right)^i = \partial^j\partial_j W^i + \frac{1}{3}\partial^i\partial_j W^j
\,.
\end{equation}
Whilst we adopt this simplification here, we expect that our method should work equally well for non-conformally flat spacetimes, and intend to develop a fully general system in future work.

Several methods may be used to solve the vector Laplacian, and these can lead to different numerical convergence properties, depending on the physical system studied. Here we summarise the two used in this work.
\begin{itemize}
\item \textit{Non-compact sources:} One approach \cite{Bowen1979GeneralFF,Bowen:1980yu} to solve this equation is based on writing the vector field $W_i$ as a sum of another vector field $V_i$ and the gradient of a scalar field $U$, 
\begin{equation}\label{eq:W_dec_1}
W_i = V_i + \partial_i U
\,,
\end{equation}
so that the vector Laplacian is expressed as
\begin{align}
\left(\bar{\Delta}_\mathrm{L}W\right)_i = \partial^j\partial_j V_i + \frac{1}{3}\partial_i\partial^j V_j + \partial^j\partial_j\partial_i U + \frac{1}{3}\partial_i\partial^j\partial_j U
\,. 
\end{align}
We have the freedom to choose $U$ in such a way that it cancels the second term in the equation above,
\begin{equation}\label{eq:vect_lapl_1}
\partial^j\partial_j U = -\frac{1}{4}\partial^j V_j~.
\end{equation}
The vector Laplacian simplifies, and the momentum constraints become three flat-space Poisson equations for $V_i$
\begin{equation}\label{eq:vect_lapl_2}
\partial^j\partial_j V_i = \frac{2}{3}\psi^6\partial_i K + 8\pi\psi^{6}S_i 
\,.
\end{equation}
A simple choice of $U$ that solves \eqref{eq:vect_lapl_1} is $\partial_i U = -V_i/4$, so that $W_i = 3V_i/4$.

\item \textit{Compact sources:}  A second approach \cite{Shibata:1999wi,Nakamura:1998qd} chooses
\begin{equation}\label{eq:compact_ansatz}
W_i = \frac{7}{8}V_i - \frac{1}{8}\left(\partial_i U + x^k \partial_i V_k\right)
\,,
\end{equation}
so that the momentum constraint yields
\begin{align}
\frac{5}{6}\partial^j\partial_j V_i - \frac{1}{6}\partial_i\partial^j\partial_j U -\frac{1}{6}x^k\partial_i\partial^j\partial_j V_k -\frac{2}{3}\psi^6\partial_i K
= 8\pi\psi^{6} S_i 
\,. 
\end{align}
If we choose $U$ such that
\begin{equation}\label{eq:vect_lapl_3}
\partial^j\partial_j U = \frac{2}{3}\psi^6 x^j\partial_j K -8\pi\psi^{6}x^jS_j 
\,,
\end{equation}
then
\begin{align}
\frac{5}{6}\partial^j\partial_j V_i - \frac{1}{6}\partial_i\partial^j\partial_j U -\frac{1}{6}x^k\partial_i\partial^j\partial_j V_k -\frac{2}{3}\psi^6\partial_i K
= 8\pi\psi^{6} S_i 
\,, 
\end{align}
which again results in the constraint becoming
\begin{equation}\label{eq:vect_lapl_4}
\partial^j\partial_j V_i = \frac{2}{3}\psi^6\partial_i K + 8\pi\psi^{6}S_i
\,.
\end{equation}
\end{itemize}
Note that in both cases we retain the term in $\partial_i K$, which means that the momentum constraints remain coupled to the Hamiltonian constraint, although in the traditional CTT approach this is usually zero. Whilst both approaches result in the same Poisson equation for $V_i$, they differ in the way in which $W_i$ and thus $\bar{A}_{ij}$ are reconstructed from the vector recovered. The latter approach results in more compact solutions and so is suited to asymptotically flat spacetimes containing compact objects, whereas the former is required for periodic spacetimes such as those used in cosmological simulations. This choice should be guided by the physical system studied, in particular it should be consistent with the required behaviour at the boundaries; a poor choice affects the convergence of the numerical algorithm. For example, extrapolating boundary conditions for $U$ and $V_i$ have provided satisfactory results in the case of black hole spacetimes.

\subsection{Solutions with black holes}
\label{sec-bhs}

Here we review solutions in the vacuum case, the so-called Bowen-York solutions of the momentum constraint, which provide initial data for boosted and rotating black holes. 

Consider vacuum solutions for which the matter source terms vanish, $\rho=S^i=0$. Assuming a moment of time symmetry, the momentum constraints are trivially satisfied by choosing $K_{ij} = K = 0$. The Hamiltonian constraint then reduces to
\begin{equation}\label{eq:hambh}
\bar{D}^2\psi=\frac{1}{8}\psi\bar{R}
\,,
\end{equation}
where $\bar{R}$ is the Ricci scalar associated to the conformal metric $\bar{\gamma}_{ij}$. Choosing the conformal metric to be flat $\bar{\gamma}_{ij}=\delta_{ij}$ makes $\bar{D}_i$ reduce to the standard flat covariant derivative so that $\bar{D}^2=\partial^i\partial_i$ and the Ricci scalar vanishes $\bar{R}=0$. The Hamiltonian constraint then becomes
\begin{equation}
\partial^i\partial_i\psi=0
\,.
\end{equation}
Spherically symmetric solutions are given by
\begin{equation}\label{eq:psi_isotropic}
\psi = 1+\frac{M}{2r}
\,,
\end{equation}
where the constant $M$ corresponds to the gravitational mass of the black hole. In addition, the solution is linear so if we want to construct multiple black hole initial data we can just superpose the single solution to obtain Brill-Lindquist initial data \cite{Brill:1963yv}
\begin{equation}
\psi = 1 + \sum_i \frac{M_i}{2r_i}
\,.
\end{equation}
The form of $\psi$ in \eqref{eq:psi_isotropic} is that of the Schwarzschild solution expressed in isotropic coordinates
\begin{equation}
dl^2=\left(1+\frac{M}{2r}\right)^4\left(dr^2+r^2\left(d\theta^2+
\sin^2\theta d\phi^2\right)\right)
\,,
\end{equation}
which can be transformed to the solution in Schwarzschild coordinates
\begin{equation}
dl^2=\left(1-\frac{2M}{R}\right)^{-1}dR^2+R^2\left(d\theta^2+
\sin^2\theta d\phi^2\right)
\,,
\end{equation}
by the following coordinate transformation
\begin{equation}
R=r\left(1+\frac{M}{2r}\right)^2
\,,
\end{equation}
where the location of the horizon for a black hole of mass $M$ is at $R=2M$ (or $r=M/2$ in isotropic coordinates).

If we want to generalise the previous solution to boosted or rotating black holes, the assumption of time symmetry must be broken which necessitates solving the momentum constraints. For the CTT case, assuming maximal slicing $K=0$ and conformal flatness, Bowen and York found the following closed form solutions \cite{Bowen:1980yu} to the constraint
\begin{equation}
\partial^j\partial_j W^i + \frac{1}{3}\partial^i\partial_j W^j = 0
\,
\end{equation}
as follows:
\begin{itemize}
\item \textit{A boosted black hole} solution given by
\begin{equation}\label{eq:boosted}
W^i = -\frac{1}{4r}\left(7\mathcal{P}^i+l^i l_j\mathcal{P}^j\right)
\,,
\end{equation}
where $l^i=x^i/r$ is the unit (coordinate) vector pointing radially outwards from the black hole and $\mathcal{P}^i$ is the linear momentum of the black hole.
\item \textit{A spinning black hole} solution for which
\begin{equation}\label{eq:spinning}
W^i = \frac{1}{r^2}\epsilon^{ijk} l_j \mathcal{S}_k
\,,
\end{equation}
where $\epsilon_{ijk}$ is the completely antisymmetric Levi-Civita tensor in three dimensions and $\mathcal{S}^i$ is the angular momentum of the black hole.
\end{itemize}
Given the linearity of the momentum constraints, we can obtain the boosted and rotating black hole solution by adding \eqref{eq:boosted} and \eqref{eq:spinning}. One therefore reconstructs the longitudinal part of the traceless extrinsic curvature tensor via \eqref{eq:Aij_L} as
\begin{align}\label{eq:bowenyork}
\bar{A}^{ij}_\mathrm{L} = \frac{3}{2r^2}\left[n^i\mathcal{P}^j + n^j\mathcal{P}^i +n_k\mathcal{P}^k\left(n^in^j-\delta^{ij}\right)\right]
- \frac{3}{r^3}\left(\epsilon_{ilk}n_j + \epsilon_{jlk}n_i\right)n^l \mathcal{S}^k
\,.
\end{align}
For such boosted and/or rotating black holes, the analytical conformal factor in \eqref{eq:psi_isotropic} no longer solves the Hamiltonian constraint and one needs to numerically solve
\begin{equation}
\bar{D}^2\psi  + \frac{1}{8}\psi^{-7} \bar{A}_{ij}\bar{A}^{ij} = 0
\,.
\end{equation}
Here the numerical approach is usually to linearise the Hamiltonian constraint by expanding around a ``guess'' solution  $\psi=\psi_0 + u$ and solve for $u$ iteratively from the linearised differential equation. In doing so one should apply the puncture method \cite{Brandt:1997tf}, which mitigates the divergences at the black hole centres.

\end{document}